\documentclass[namedreferences]{solarphysics}

\usepackage[hyperref,optionalrh,showbiblabels]{spr-sola-addons} 
\usepackage{graphicx}        
\usepackage{color}           
\usepackage{breakurl}        

\chardef\us=`\_

\begin{document}

\begin{article}
\begin{opening}

   \title{Meridional Motion and Reynolds Stress from Debrecen Photoheliographic Data}
\author[addressref=aff1,corref,email={davor.sudar@gmail.com}]{\inits{D.}\fnm{Davor}~\lnm{Sudar}~\orcid{0000-0002-1196-6340}}
\author[addressref=aff1,email={romanb@geof.hr}]{\inits{R.}\fnm{Roman}~\lnm{Braj\v{s}a}}
\author[addressref=aff2,email={ivica.skokic@gmail.com}]{\inits{I.}\fnm{Ivica}~\lnm{Skoki\'{c}}}
\author[addressref=aff3,email={ipoljancic@phy.uniri.hr}]{\inits{I.}\fnm{Ivana}~\lnm{Poljan\v{c}i\'{c} Beljan}}
\author[addressref=aff4,email={hw@leibniz-kis.de}]{\inits{H.}\fnm{Hubertus}~\lnm{W\"{o}hl}}

\address[id=aff1]{Hvar Observatory, Faculty of Geodesy, Ka\v{c}i\'{c}eva 26, University of Zagreb, 10000 Zagreb, Croatia}
\address[id=aff2]{Astronomical Institute of the Czech Academy of Sciences, Fri\v{c}ova 298, 251 65 Ond\v{r}ejov, Czech Republic}
\address[id=aff3]{Department of Physics, University of Rijeka, Radmile Matej\v{c}i\'{c} 2, 51000 Rijeka, Croatia}
\address[id=aff4]{Kiepenheuer-Institut f\"{u}r Sonnenphysik, Sch\"{o}neckstr. 6, 79104 Freiburg, Germany}

\runningauthor{D. Sudar {\it et al.}}
\runningtitle{Meridional Motion and Reynolds Stress}

\begin{abstract}
The Debrecen Photoheliographic Data catalogue is a continuation of the Greenwich Photoheliographic Results providing daily positions
of sunspots and sunspot groups.
We analyse the data for sunspot groups focusing on meridional motions and transfer of angular momentum towards the solar equator.
Velocities are calculated with a daily shift method including an automatic iterative process of removing the outliers.
Apart from the standard differential rotation profile, we find meridional motion directed towards the zone of solar activity.
The difference in measured meridional flow in comparison to Doppler measurements and some other tracer measurements
is interpreted as a consequence of different flow patterns inside and outside of active regions. We also find a 
statistically significant dependence of meridional motion on rotation velocity residuals confirming the transfer of angular momentum
towards the equator. Analysis of horizontal Reynolds stress reveals that the transfer of angular momentum is stronger
with increasing latitude up to about 40$^{\circ}$ where there is a possible maximum in absolute value.
\end{abstract}

\end{opening}


\section{Introduction}
     \label{S-Introduction} 
The Debrecen Photoheliographic Data (DPD) catalogue was started as a continuation of the Greenwich Photographic Results (GPR).
Royal Greenwich Observatory ceased its photoheliographic program in 1977 and the International Astronomical Union commissioned Debrecen
Observatory, Hungary, to continue the project \citep{Wayman1980}. The DPD catalogue is also supplemented by solar images from other observatories
(for details see \citet{Baranyi2016}) to fill in the
gaps in Debrecen observations.

The GPR dataset represents one of the most valuable resources in studying behaviour of the Sun over long time periods. It comes as no surprise
that various portions of the dataset were used in a large number of papers studying solar rotation \citep{Newton1951, Ward1965, Ward1966, Balthasar1980,
Arevalo1982, Balthasar1986b, Brajsa2002, Brajsa2004, Ruzdjak2004, Ruzdjak2005}. Often the dataset was extended into the future by using
observations by the Solar Observing Optical Network of the United States Air Force/National Oceanic and Atmospheric Administration (SOON/USAF/\-NOAA)
\citep{Pulkkinen1998a, Javaraiah2003, Zuccarello2003, Javaraiah2005, Brajsa2007, Javaraiah2010, Sudar2014}. \citet{Pulkkinen1998a} also extended
the analysis into the past by using observations made by
Carrington and Sp\"{o}rer. Observations of sunspots from other observatories were also used: {\it e.g.}, Kanzel\-h\"{o}he \citep{Lustig1983}, Mt. Wilson \citep{Gilman1984},
Mitaka \citep{Kambry1990}, Kodaikanal \citep{Gupta1999}, and Abastumani \citep{Khutsishvili2002}.

The transport of angular momentum towards the solar equator, necessary for maintaining the observed solar differential
rotation profile, is often attributed to Reynolds stresses which result from mutual dependence
between meridional motions and rotation velocity residuals \citep{Ruediger1980, Canuto1994, Pulkkinen1998b}.
Motions of only a few m s$^{-1}$ in both velocity components are sufficient to generate horizontal Reynolds stress of the order of
several 10$^{3}$ m$^{2}$ s$^{-2}$ which is sufficient to maintain the observed solar rotation profile \citep[][and references therein]{Schroter1985}.
Actual observations show the required value of several 10$^{3}$ m$^{2}$ s$^{-2}$ for horizontal Reynolds stress
\citep{Ward1965, Belvedere1976, Schroter1976, Gilman1984, Pulkkinen1998b, Vrsnak2003,
Sudar2014}. \citet{Ward1965}, \citet{Gilman1984}, \citet{Pulkkinen1998b}, \citet{Vrsnak2003}, \citet{Sudar2014} also showed that the absolute value
of horizontal Reynolds stress increases
with latitude up to about 30$^{\circ}$ where there is a possible maximum.

Velocity components of the Reynolds stress (meridional motions and rotation velocity residuals) are also a subject of investigation.
\citet{Howard1980} found a cyclic pattern of alternating faster and slower rotation bands with a period of $\approx$11 years. Further
confirmation of these torsional oscillations was reported by \citet{Ulrich1988}, \citet{Howe2000}, \citet{Haber2002}, \citet{Basu2003}. Recently, \citet{Sudar2014}
tried to find a similar pattern in sunspot groups velocity data, but were unable to detect such a signal when folding all solar cycles
to one phase diagram. Doppler measurements show meridional flow directed towards solar poles on all latitudes \citep{Duvall1979, Howard1979,
Hathaway1996, Zhao2004, Kosovichev2016}. Most of the time the flow was of the order of 20 m s$^{-1}$, but episodes of significantly
larger flow have also been detected \citep{Hathaway1996}. Sometimes almost no meridional flow \citep{Lustig1990} or even opposite
flow \citep{Perez1981} was detected in Doppler data. Various tracers, such as sunspots, sunspot groups, small magnetic features, coronal
bright points (CBPs) and solar plages, were also used to measure the meridional motion. Sunspots and sunspot groups were most frequently used
and most of the results show flow outward from the centre of the solar activity \citep{Tuominen1984, Howard1986, Howard1991b, Kambry1991, Woehl2001}.
Recently, however, \citet{Sudar2014} reported the flow which is almost exactly the opposite. Similar flow to \citet{Sudar2014} was found by
\citet{Howard1991a} by tracing solar plages. \citet{Komm1993} and \citet{Snodgrass1996} used small magnetic features as tracers and found
similar, but still in some aspects mutually different, results. \citet{Komm1993} found poleward flow at all latitudes with a maximum amplitude of
about 10 m s$^{-1}$, while \citet{Snodgrass1996} obtained a flow which is directed out of the centre of solar activity with poleward
flow at higher latitudes with similar amplitude to that of \citet{Komm1993}. Analysing the motion of CBPs, \citet{Sudar2016} detected poleward meridional
flow for all latitudes with an amplitude of about 30 m s$^{-1}$.

The main focus of this paper is to analyse the meridional motion revealed by tracing sunspot groups over the solar surface with DPD which
has never been used before for this purpose. We also attempt to explain the wide variety of results for meridional flow in Section~\ref{Sect_Discussion}
in a consistent way. Another very important aspect of this work is to confirm that Reynolds stress is the main generator of the observed
solar rotation profile as already suggested by a number of previous theoretical and empirical papers.

\section{Data and Reduction Methods}
Our dataset is created from positions and times of sunspot groups measured in DPD\footnote{see \url{http://fenyi.solarobs.csfk.mta.hu/en/databases/DPD/}.}
\citep{Baranyi2016, Gyori2017} during a period from 1974 to 2016.
Positions of sunspot groups are typically determined once per day. We use the daily change in position of each sunspot group
to calculate both meridional and rotational speeds:
\begin{eqnarray}
\omega_{\rm rot} &=& \frac{\Delta \rm CMD}{\Delta t},\\ 
\omega_{\rm mer} &=& \frac{\Delta b}{\Delta t},
\end{eqnarray}
where $\Delta \rm CMD$ is the difference between central meridional distances (CMD)
and $\Delta b$ is the difference in latitude of two consecutive positions of the sunspot group
in time $\Delta t$ (usually 1 day).

This results in 59090 data points for each speed. Rotation speeds are transformed
from synodic to sidereal values \citep{Skokic2014}.
Assigning particular latitude, $b$, to each velocity is not as straightforward as it seems. With two measurements of
position to calculate one velocity, we could use the latitude of the first or the second measurement. Even average latitude
seems like a possible candidate. However, as \citet{Olemskoy2005} pointed out, an uneven distribution of sunspots (or any
other kind of tracer) in latitude creates a problem with some of the possible choices. They have shown that the gradient
of the latitudinal distribution can create false meridional flows if the latitude of the second measurement is used.
The simplest way to avoid this problem is to use the latitude of the first measurement and this is the choice we make
in this and previous papers \citep{Sudar2014, Sudar2015, Sudar2016}.

\begin{figure}
\centerline{\includegraphics[width=0.75\textwidth,clip=]{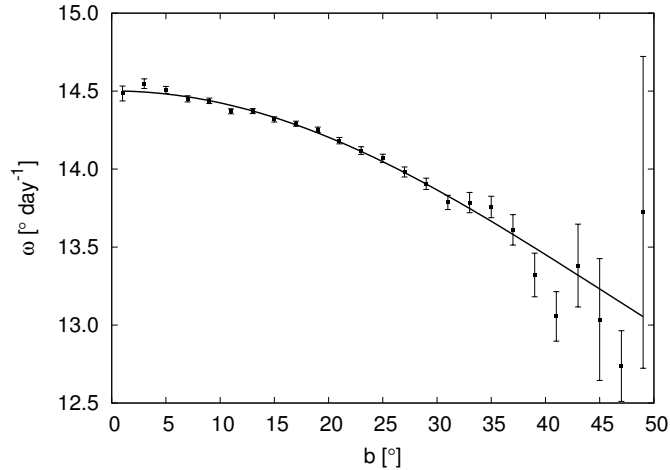}}
\caption{Best fit solar differential rotation profile, $\omega(b)$, is shown with the solid line. Solid squares
with error bars are bin averaged values of the rotation velocity where bins are 2$^{\circ}$ wide in latitude.}
\label{Fig_Rotation}
\end{figure} 

Determination of the position of a sunspot on the limb can introduce large errors due to projections effects. In addition,
\citet{Arevalo1982} found that the calculated rotation profile changes depending on the cut-off longitude.
\citet{Woehl1983} also reported that the determined rotation velocities are larger for smaller cut-off longitudes.
Such changes were attributed to Wilson depression by \citet{Balthasar1983}. Therefore we limit the dataset to
$\pm$58$^{\circ}$ in CMD which alleviates these problems \citep{Stark1981, Balthasar1986a}.
Nevertheless, outliers resulting from misidentification of sunspot groups in subsequent measurements or other
errors are still present, so we use an iterative filtering method similar to the one used in \citet{Sudar2016}.
In the first step we calculate the solar rotation profile:
\begin{equation}
\label{Eq_rotProfile}
\omega(b)=A + B\sin^{2}(b),
\end{equation}
where $b$ is the latitude, and then we calculate the rotation velocity residuals by
subtracting the individual rotational speeds from the average rotation profile given by Equation~(\ref{Eq_rotProfile}).
In the next step we calculate the lower quartile, $Q_{1}$, and upper quartile, $Q_{3}$,
for rotation velocity residuals and finally remove the so-called hard outliers which lie outside of the range:
\begin{equation}
\label{Eq_interquartile}
[Q_{1} - k(Q_{3} - Q_{1}), Q_{3}+k(Q_{3} - Q_{1})],
\end{equation}
where we chose $k$=3.5. Since outliers could potentially influence the calculated
rotation profile, we use the remaining dataset to recalculate the solar rotation profile.
The new rotation profile changes the values of rotation velocity residuals which then
need to be checked for outliers. This process continues iteratively
until no data points are outside of the interquartile range (Equation~(\ref{Eq_interquartile})).
In the end we obtain a dataset consistent with the calculated rotation profile and
with outliers removed.
In each iteration we also remove the outliers in meridional velocity using the same form of the interquartile
range. The process converges very fast and only four iterations were necessary in our case.
Outlier limits turn out to be $\pm$4.5 $^{\circ}$ day$^{-1}$ and
$\pm$2.1 $^{\circ}$ day$^{-1}$ for rotation velocity residuals and meridional velocities, respectively.

After removing the outliers we end up with 53283 data points. The best fit rotation profile is shown in Figure~\ref{Fig_Rotation}
where the coefficients of the profile (Equation~(\ref{Eq_rotProfile})) are $A$ = 14.5011$\pm$0.0081 $^{\circ}$ day$^{-1}$ and
$B$ = -2.540$\pm$0.073 $^{\circ}$ day$^{-1}$. In the same figure we also show average values of $\omega(b)$ in 2$^{\circ}$
wide bins in latitude. The error bars shown in Figure~\ref{Fig_Rotation} become fairly large for $b> 35^{\circ}$,
which is a consequence of the fact that sunspots rarely appear above this latitude. In terms of expansion in Gegenbauer polynomials
\citep{Snodgrass1985} the rotation profile coefficient become $A_{\rm G}=13.993$ $^{\circ}$ day$^{-1}$ and $B_{\rm G}=-0.51$ $^{\circ}$ day$^{-1}$.

In the rest of the paper we use only residual rotation velocities and meridional velocities which are
transformed from angular values to their linear counterparts taking into account the latitude of the first
measurement. The conversion factor, with $R_{\odot}$=6.96$\cdot$10$^{8}$ m, is 140.6 m s$^{-1}$ day $({^{\circ}})^{-1}$, while rotation velocity residuals
are additionally multiplied with the cosine of latitude.
Additionally, meridional speeds are symmetrized, so that negative value of meridional speed reflects motion towards
the equator for both solar hemispheres. This is achieved by transforming calculated meridional speeds with
$v_{\rm mer}=-\partial b/\partial t$ for the southern solar hemisphere, where we assign negative values of $b$
for southern latitudes. 

\section{Results}

\begin{figure}
\centerline{\includegraphics[width=0.75\textwidth,clip=]{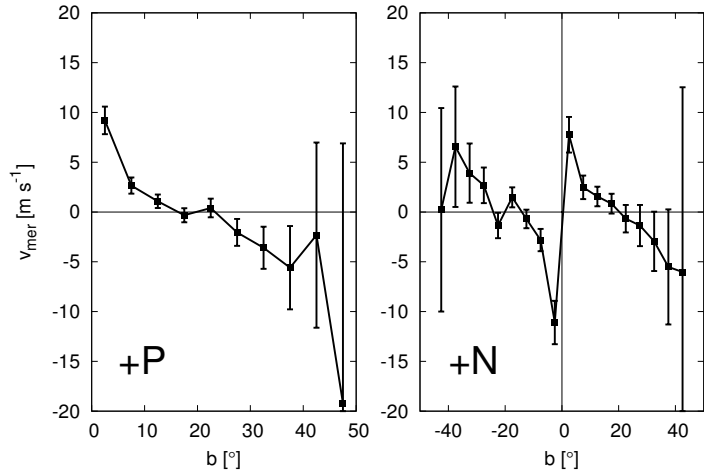}}
\caption{Meridional flow, $v_{\rm mer}$ as a function of latitude, $b$, is shown with black squares with error bars.
Averaging bins are 5$^{\circ}$ wide in latitude. On the left hand side we show all data folded into one hemisphere where
meridional velocity is asymmetrically transformed for southern latitudes. In this part of the plot positive values indicate
motion towards the poles (+P). On the right hand side of the plot meridional velocities are not transformed and we show
both hemispheres. Positive values indicate motion towards the northern solar pole (+N).}
\label{Fig_Meridional}
\end{figure} 

In Figure~\ref{Fig_Meridional} we show the dependence of the meridional flow on latitude, $b$.
Black squares with error bars depict average meridional flow in 5$^{\circ}$ bins of latitude.
We can see that for $-15^{\circ}< b<15^{\circ}$ meridional flow is towards the solar poles.
In the right hand side of the plot we see that at the equator meridional flow is $\approx$0 m s$^{-1}$
and that the flow is asymmetrical around the equator.
At mid-latitudes meridional flow becomes zero again on both hemispheres and probably turns to flow towards the equator
at even higher latitudes. Similar behaviour was found by \citet{Sudar2014} who also concluded
that the latitude at which the flow becomes zero is the centre of solar activity defined
by the latitudinal distribution of sunspots for each phase of the solar activity cycle.

On the other hand, by using Doppler line shifts, most researchers found
poleward flow of about 20 m s$^{-1}$ for all latitudes \citep{Duvall1979, Howard1979,
Hathaway1996}. Similar values were found by using high-resolution magnetograms
\citep{Komm1993} and by applying time–-distance helioseismology \citep{Zhao2004}.
Moreover, by using coronal bright points (CBP) as tracers, \citet{Sudar2016} found
poleward meridional flows everywhere except at the equator where the flow was zero.
This discrepancy in meridional flow between sunspot measurements and other observations is
discussed in the next section.

\begin{figure}
\centerline{\includegraphics[width=0.75\textwidth,clip=]{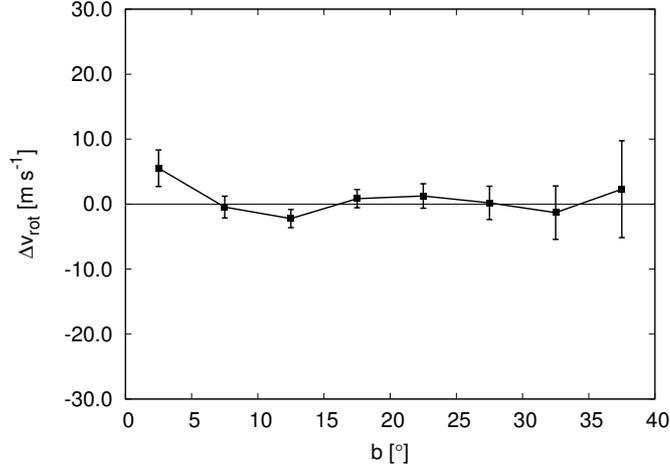}}
\caption{Rotation velocity residuals, $\Delta v_{\rm rot}$, as a function of latitude, $b$, are
shown with black squares with error bars. Width of averaging bins in latitude is 5$^{\circ}$.
Positive values show faster than average rotation.}
\label{Fig_Torsional}
\end{figure}
Average values of the rotation velocity residuals, $\Delta v_{\rm rot}$, in bins of 5$^{\circ}$ in latitude, $b$,
are shown in Figure~\ref{Fig_Torsional} with solid squares with error bars. 
No significant deviation from $\Delta v_{\rm rot}$=0 can be detected.

\begin{figure}
\centerline{\includegraphics[width=0.75\textwidth,clip=]{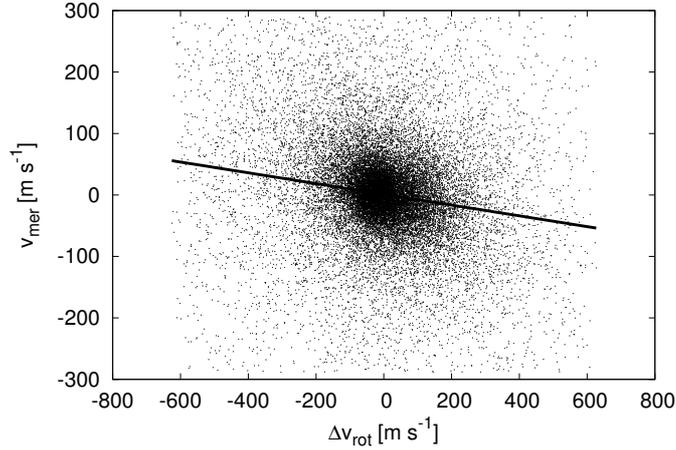}}
\caption{Individual observations are shown with dots in the $v_{\rm mer}$ -- $\Delta v_{\rm rot}$ parameter space.
We also show the best linear fit function (Equation~(\ref{Eq_v_correlation})) with a solid line.}
\label{Fig_v_correlation}
\end{figure} 
One important feature of the solar rotation profile is that lower latitudes rotate
faster than higher latitudes which implies that most of the angular momentum of the
Sun is located in the zones around the equator. Therefore, it is reasonable to assume
that there must be some mechanism which transfers angular momentum towards lower latitudes.
By studying the relationship between meridional velocities, $v_{\rm mer}$, and rotation velocity
residuals, $\Delta v_{\rm rot}$, we can directly observe if this is what is really happening.
The relationship between the two velocities is shown in Figure~\ref{Fig_v_correlation}. We also
show the best linear fit to the data with the solid line, obtaining the following relation:
\begin{equation}
\label{Eq_v_correlation}
v_{\rm mer} = (-0.0876\pm 0.0021)\Delta v_{\rm rot} + (1.01\pm 0.34)\mathrm{m\ s^{-1}}.
\end{equation}
The slope in Equation~(\ref{Eq_v_correlation}) is statistically significant when
compared with its uncertainty (relative error $\approx$2.4\%) which shows that on average the two values
are not independent. The fact that the slope is negative shows that on average the
angular momentum is indeed transferred from higher to lower latitudes.
In a previous study, using the sunspot groups data from the GPR and the SOON/USAF/NOAA obtained in the period
1878–-2011, \citet{Sudar2014} found a very similar value for the slope $(-0.080\pm 0.002)$.

\begin{figure}
\centerline{\includegraphics[width=0.75\textwidth,clip=]{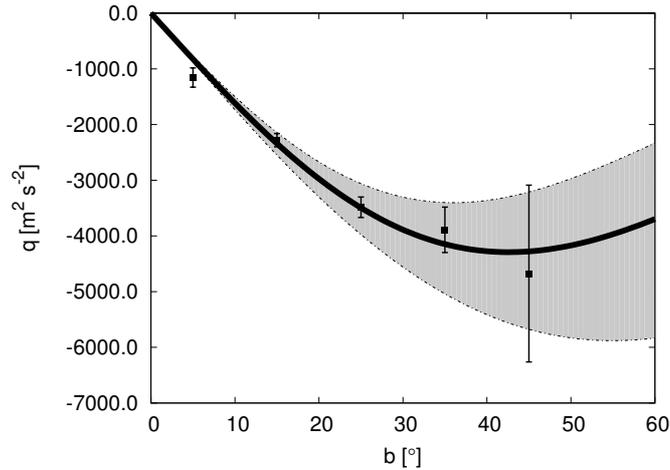}}
\caption{Reynolds stress, $q=<v_{\rm mer}\Delta v_{\rm rot}>$, is shown as a function of latitude with black squares with
error bars. Averaging bins are 10$^{\circ}$ wide in latitude. We also show the best fit function of the form given
in Equation~(\ref{Eq_ModelE}) with best fit coefficients from Table~\ref{Tab_coeffs}. Shaded area depicts the range of
possible values defined by the errors of the best fit coefficients (Table~\ref{Tab_coeffs}).}
\label{Fig_Reynolds}
\end{figure} 
Reynolds stress is thought to be the
main generator of the differential rotation on the Sun \citep{Ruediger1980, Pulkkinen1998b} which works against
the diffusive decay.
Covariance of meridional velocities and rotation velocity residuals, $q=<\Delta v_{\rm rot}v_{\rm mer}>$,
is the horizontal component of the Reynolds stress tensor and in Figure~\ref{Fig_Reynolds} we show
its dependence on latitude. Values of $q$ in 10$^{\circ}$ bins in latitude, $b$, are shown as solid squares
with error bars. We can see that the average value of $q$ is negative for all latitudes meaning that
there is a net angular momentum transfer towards lower latitudes. This picture might seem to be in conflict
with Figures~\ref{Fig_Meridional} and \ref{Fig_Torsional} which show average behaviour of $v_{\rm mer}$ and $\Delta v_{\rm rot}$
with respect to latitude, $b$. However, we must remember that the average of the product is not equal to the
product of averages, i.e. $q=<v_{\rm mer}\Delta v_{\rm rot}> \neq <v_{\rm mer}><\Delta v_{\rm rot}>$. Negative values of $q$
actually imply that $v_{\rm mer}$ and $\Delta v_{\rm rot}$ are not mutually independent variables and that the relationship
between them must be similar to the one given in Equation~(\ref{Eq_v_correlation}).

\begin{table}
\caption{ Table of the best fit coefficients (Equation~(\ref{Eq_ModelE})).}
\label{Tab_coeffs}
\begin{tabular}{lcc}
  \hline
Coef & Value & Rel. Error\\
  \hline
$c_{1}$ [m$^2$ s$^{-2}$ $(^{\circ})^{-1}$] & $-166 \pm 10$ & 6.2\%\\
$c_{3}$ [$(^{\circ})^{-2}$] & $0.00028 \pm 0.00011$ & 39.9\%\\
  \hline
\end{tabular}
\end{table}
\citet{Sudar2014} introduced an empirical exponential cut-off function which describes the decreasing
trend of $q(b)$ from the equator to higher latitudes and allows for a possible minimum. In this work we simplify
the form of this relationship by requiring that the function is perfectly asymmetric around zero ($q(b)=-q(-b)$) leading to
the expression:
\begin{equation}
\label{Eq_ModelE}
q = c_{1}b {\rm e}^{-c_{3}b^{2}}.
\end{equation}
This choice is justified by the appearance of function $q$ in Figure 10 in
\citet{Sudar2014}, Figure 15 in \citet{Canuto1994}, and also by the fact that \citet{Sudar2014} found that the value of parameter $e_{2}$
in their model was $69\pm80$ m s$^{-2}$. The function is plotted in Figure~\ref{Fig_Reynolds} with a thick solid line. In the same plot
we also shade the area defined by errors of the coefficients $c_{1}$ and $c_{3}$ (Table~\ref{Tab_coeffs}).
Using the errors of the coefficients we can also calculate the depth and location of the minimum in $q(b)$ plot with their
respective errors by using the method of error propagation. The minimum is thus located at $b_{\rm min}=(42.3\pm 8.3)^{\circ}$ with a value
of $q(b_{\rm min})=(4250\pm880)$ m$^{2}$ s$^{-2}$.
A similar result was obtained by \citet{Sudar2014} who suggested that the $q(b)$ relationship has a minimum ($q\approx -3000$ m s$^{-2}$) around 25-30$^{\circ}$.
We don't consider the difference between the two results to be significant, especially considering the scarcity of sunspots at higher
latitudes where the minimum is supposed to be.

\section{Discussion}
\label{Sect_Discussion}
The rotation profile calculated in this work is almost identical to the result obtained by \citet{Sudar2014} who also analysed the rotation
by tracing sunspot groups albeit for a different time period and different observing stations. So, the DPD series proves to be a rather good continuation of the GPR
data set. For a more thorough comparison of
solar rotation profiles obtained by different methods we refer the reader to \citet{Woehl2010} and \citet{Sudar2015}.

In Figure~\ref{Fig_Torsional} we show the average of rotation velocity residuals, $\Delta v_{\rm rot}$, as a function of latitude, $b$.
This plot should not be confused with the torsional oscillation pattern, because torsional oscillations show deviations from the mean velocity
in time. When we average these variations over a long time, as we do in Figure~\ref{Fig_Torsional}, such a plot is more indicative
of the quality of the fit of the solar rotation profile function (Equation~(\ref{Eq_rotProfile})). In this context, we can say that, apart
from the bin at $b=2.5^{\circ}$, the average value of $\Delta v_{\rm rot}$ is zero. We are not sure if the slight discrepancy from this rule at $b=2.5^{\circ}$
is of any significance.

The most interesting are the results for the meridional flow (Figure~\ref{Fig_Meridional}) because they are the most controversial. A quote
by \citet{Hathaway1996} is perhaps the most illustrative of the problem: ``Unfortunately, previous measurements of the Sun's
meridional circulation have produced a bewildering array of results that appear to be of no help at all in constraining theory."
On one side we have sunspot measurements (and other features closely associated with sunspots such as plages) which show opposite flows
on opposite sides of the centre of the solar activity and on the other hand we have Doppler measurements which predominately show poleward
motion for all latitudes, regardless of the centre of the solar activity. There are also tracers, such as CBPs \citep{Sudar2016},
which also show average poleward motion everywhere of approximately the same amplitude as Doppler measurements. Small magnetic features
analysed with different methods by \citet{Komm1993} and \citet{Snodgrass1996} showed different behaviour around the centre of
solar activity for meridional motion.

So, if we measure the same phenomenon with different techniques, how can we consistently interpret these differing results? Firstly there is an internal problem
with tracer measurements where some authors \citep{Tuominen1984, Howard1986, Howard1991b, Kambry1991, Woehl2001}
reported a flow out of the centre of activity, while others detected a flow towards the
centre of activity \citep{Howard1991a, Sudar2014}. This difference can probably be explained by an easy-to-make error, assuming that it
is irrelevant to which latitude one assigns the observed meridional velocity. \citet{Olemskoy2005} demonstrated and \citet{Sudar2014} later
verified that it is necessary to assign the latitude of the first measurements due to the uneven distribution of tracers in latitude. Otherwise one would
obtain almost exactly opposite and false flow. Technically it is possible to use the latitude of the last measurement of position, but the
velocity would have to be weighted taking into account the frequency of tracers at the starting latitude. Since the tracer distribution can also vary
during the solar cycle, the second method turns out to be very complicated compared to the neat trick of assigning the latitude of the first measurement.
As a consequence we take the results by \citet{Sudar2014} as a reference model for sunspot measurements which are also confirmed by the results for
meridional flow in this paper (Figure~\ref{Fig_Meridional}). However, we must point out that \citet{Sivaraman2010}, who also used first latitude
as a reference point, found a flow which is directed towards the solar equator for all latitudes. What we find intriguing in their results is that it appears
that meridional flow is not zero at the equator, which is especially noticeable for the Mt. Wilson data they used. On the right hand side of our Figure~\ref{Fig_Meridional}
we can see that the average meridional motion is $\approx$0 which makes the motion clearly asymmetrical on the two solar hemispheres.

Small magnetic features were also used as tracers by past authors and although their latitudinal distribution is not the same as the one for sunspots we encounter
the same problem. Their distribution overall falls from the equator towards the poles, but it also has peaks around centres of solar activity
\citep{Harvey1993}.
Such a distribution gradient can create false flows out of the centre of activity in a similar manner as with sunspots, if proper care of
assigning latitude was not taken. In this context we can explain the results by \citet{Snodgrass1996}, but the meridional flow measured
by \citet{Komm1993} still requires some explanation. First we note that the amplitude of meridional motion in their work was about 10 m s$^{-1}$
which is substantially lower then most Doppler measurements and analysis of CBP data, so it is possible that their result was also influenced
just by the overall drop in distribution of tracers towards the poles where the velocity resolution was not sufficient to be influenced by
the peaks near centres of activity. Of course, it is also possible that they measured the real flow and that the fairly low amplitude they
obtained is consistent with the wide range of possible variations mentioned by \citet{Hathaway1996}.

Analysing the flow in the upper convective zone with time-distance helioseismology \citet{Zhao2004} found mostly poleward meridional flow (their Figure~3a).
However, when they subtracted the flow from Carrington Rotation 1911 which occurred during the solar minimum year 1996 they obtained the {\em residual}
meridional flow (their Figure~3b) which in all important aspects looks the same as the meridional flow we see in this work (Figure~\ref{Fig_Meridional}) and
in Figure~3 in \citet{Sudar2014}. Applying ring-diagram analysis to GONG Dopplergrams \citet{Gonzales2008} also found residual meridional flows
very similar to our meridional flow (their Figures.~4 and 5). Moreover, map plots of residual meridional motion in \citet{Gonzales2008} (their Figure~6)
and \citet{Gonzales2010} (their Figure~3) look a lot like a similar map plot for meridional motion of sunspot group data in \citet{Sudar2014} (their Figure~2).

The key in understanding why residual meridional flow in Doppler measurements looks like total meridional motion in sunspot groups measurements is in
appreciating that sunspot groups do not cover all of the solar surface, but are limited to active regions. When \citet{Zhao2004} subtracted the
meridional flow measured during the solar cycle minimum when active regions rarely appear they obtained the residual flow dominated by the motion in and around
active regions and that is the area where sunspot groups are found. Therefore, it is perfectly understandable why \citet{Zhao2004} found that
their residual meridional flows in the period from 1997-2002 converge towards activity belts with the magnitude of 2-8 m s$^{-1}$ in both solar hemispheres just as we see in our
total meridional flow (Figure~\ref{Fig_Meridional}). By analysing sunspot group data we get no information about the dominant poleward meridional flow outside of the active regions.
Even the analysis of CBP data \citep{Sudar2016} supports this hypothesis, because detection of CBPs with the segmentation algorithm above bright active
regions is almost impossible. As a result, \citet{Sudar2016} found the dominant poleward meridional motion originating outside of the active regions.

An alternative to our hypothesis is that the anchor depth of magnetic features plays a role in the meridional flow pattern. Anchor depth
has been used as an explanation for differences in rotation rates measured by different tracers and different observing methods.
An useful overview of this problem is given in \citet{Beck2000}. The rotation rate of surface features are usually connected to depth
corresponding to the rotation rate obtained with helioseismology. In general larger features are associated with larger depths, but
there are exceptions like for example recurrent sunspot groups. It is logical to assume that the meridional flow of surface features also depicts
the motion of deeper layers which is what \citet{Sivaraman2010} concluded based on observing different size sunspot groups. Further work on meridional
flow at different depths, possibly in connection with the phase of the solar cycle, is still necessary to shed more light on the subject.

Another interesting question is how does this difference between sunspot group data and CBP data affect the observed horizontal Reynolds stress. Sunspot group
measurements \citep{Ward1965, Gilman1984, Sudar2014} show a fairly large negative value of $q\approx -3000$ m$^{2}$ s$^{-2}$ located near 30$^{\circ}$ in latitude. A slightly
larger value at higher latitude is also found in this work. \citet{Canuto1994} already confirmed that this result is in agreement with the theoretical curve using
only standard values for parameters in solar conditions.
On the other hand, analysis of CBP data by \citet{Sudar2016} yields a highly uncertain value of only $q\approx -1500$ m$^{2}$ s$^{-2}$ around 25$^{\circ}$. The time period
covered in \citet{Sudar2016} is less than 6 months in the rising phase of the solar cycle, but we don't think that the phase of the solar cycle is the cause of
such a low value of $q$ in CBP data. \citet{Sudar2014} showed a map of $q$ versus latitude and phase of the solar cycle and found that the locations
and depth of the horizontal Reynolds stress, $q$, are persistent at least from the beginning of the solar cycle until well past the solar activity maximum.
In the declining phase of the solar cycle, the latitudinal distribution of sunspots prevents making a firm conclusion about the behaviour of $q$. So, we suggest that
the horizontal Reynolds stress is stronger in and around active regions which means that most of the angular momentum transfer towards the
equator takes place in these areas. Another possibility is that the height of the tracers also plays a role, or in other words that Reynolds stress is
less pronounced in the layers above the photosphere. This would also imply that CBPs are not firmly rooted to the photosphere and we don't consider
this to be very likely.

\section{Conclusion}
We analyse the motion of sunspot groups obtained by Debrecen Observatory in the period 1974--2016.
The calculated solar rotation profile is in agreement with other authors especially those investigating also sunspot groups.
This shows that the DPD catalogue is an adequate successor of the GPR dataset.

The observed meridional motion is typical for the sunspot groups, but different than Doppler and CBP measurements.
We suggest that the difference lies in the fact that sunspot groups are located within active regions where the meridional motion
is different than outside of those areas. This idea is supported by {\em residual} meridional flow obtained by subtracting the
dominant poleward meridional flow in Doppler data which then looks almost the same as the total meridional flow we see
when analysing sunspot group data. \citet{Canuto1994} derived a model of Reynolds stress which is driven by buoyancy
which acts as a source of convective turbulence.
If we take into account that in strong magnetic fields within sunspots convection is inhibited, it might be possible
that the velocity components of the horizontal Reynolds stress are also affected and, therefore, show different behaviour
in and outside of active regions.

The horizontal Reynolds stress calculated in this work confirms the results in all important aspects obtained by \citet{Sudar2014} who used data from different
observatories. The latitudinal behaviour of the horizontal Reynolds stress shows the transport of angular momentum towards the equator
everywhere, increasing in absolute value from the equator to a possible maximum value somewhere around 25--40$^{\circ}$ which is also consistent with
earlier studies based on sunspot data \citep{Ward1965, Belvedere1976, Schroter1976, Gilman1984, Pulkkinen1998b}.
By comparing the horizontal Reynolds stress obtained with sunspot group data and CBP data we suggest that larger values of
horizontal Reynolds stress, and consequently angular momentum transfer, are found within active regions. Analysis of meridional
motion and horizontal Reynolds stress values with CBP data as a function of proximity to active regions might prove or disprove
this hypothesis.

Measurements of the latitudinal dependence of the covariance of rotation velocity residuals and meridional flow presented in this work strongly suggest that
the Reynolds stress is the dominant mechanism which explains the observed characteristics of the solar differential rotation.
The observed horizontal Reynolds stress is consistent with the transport of angular momentum towards the solar equator both
qualitatively (correct sign) and quantitatively (absolute value). On the other hand, the model of axisymmetric meridional circulation,
which could also produce transfer of the angular momentum towards the equator, is strongly disfavored because it requires
surface meridional motions in the direction towards the solar equator \citep{Gilman1981, Stix1989b, Foukal2013} which contradicts the observations
in the present and other papers.

\begin{acks}
The authors wish to thank the staff of Debrecen Observatory for maintaining and organising the
DPD catalogue.
This work has been supported in part by the Croatian
Science Foundation under the project 6212 ``Solar and Stellar Variability".
It has also received funding from the SOLARNET project (312495, 2013-2017) which is an Integrated Infrastructure
Initiative (I3) supported by FP7 Capacities Programme.
\end{acks} 

\section*{Disclosure of Potential Conflicts of Interest}
The authors declare that they have no conflicts of interest.

\bibliographystyle{spr-mp-sola}
\bibliography{DebrecenRot} 

\end{article} 

\end{document}